\documentclass[useAMS,usenatbib,usegraphicx]{mn2e}
\usepackage{times,amssymb}
\title[{\it Kepler} results on compact pulsators V: sdBV stars in binaries]
{First {\it Kepler} results on compact pulsators V: Slowly pulsating subdwarf B stars in short-period binaries}

\author[S. D. Kawaler et al.]
{S. D. Kawaler$^1$,
M. D. Reed$^2$,
R. H. {\O}stensen$^3$,
S. Bloemen$^3$,
D. W. Kurtz$^4$,
A. C. Quint$^2$,
\newauthor
R. Silvotti$^5$,
A. S. Baran$^{1,6}$,
E. M. Green$^7$,
S. Charpinet$^8$,
J. Telting$^9$,
C. Aerts$^3$,
\newauthor
G. Handler$^{10}$,
H. Kjeldsen$^{11}$,
J. Christensen-Dalsgaard$^{11}$,
W. J. Borucki$^{12}$,
D. G. Koch$^{12}$
\and
J. Robinson$^{13}$\\
$^1$Department of Physics and Astronomy, Iowa State University, Ames, IA 50011 USA\\
$^2$Department of Physics, Astronomy and Materials Science,
 Missouri State University, 901 S. National, Springfield, MO 65897 USA\\
$^3$Instituut voor Sterrenkunde, K.U.~Leuven, Celestijnenlaan 200D, 3001 Leuven, Belgium\\
$^4$Jeremiah Horrocks Institute of Astrophysics, University of Central Lancashire, Preston, PR1 2HE, UK\\
$^5$INAF-Osservatorio Astronomico di Torino, Strada dell'Osservatorio 20, 10025 Pino Torinese, Italy\\
$^6$Krakow Pedagogical University, ul. Podchor\c{a}\.{z}ych 2,30-084 Krak\'{o}w, Poland\\
$^7$Steward Observatory, University of Arizona, Tucson, AZ 85721, USA\\
$^8$Laboratoire d'Astrophysique de Toulouse-Tarbes, Universit\'e de Toulouse, CNRS, 14 Av. E. Belin, 31400 Toulouse, France\\
$^9$Nordic Optical Telescope, 38700 Santa Cruze de La Palma, Spain\\
$^{10}$Institute f\"ur Astronomie, Universit\"at Wien, T\"urkenschanzstrasse 17, 1180 Wien, Austria\\
$^{11}$Department of Physics and Astronomy, Aarhus University, DK-8000 Aarhus C, Denmark\\
$^{12}$NASA Ames Research Center, MS 244-30, Moffett Field, CA  94035, USA\\
$^{13}$Alpha Control, USSC, Houston, TX, USA
}

\begin{document}

\date{accepted 3 August 2010}
\pagerange{\pageref{firstpage}--\pageref{lastpage}} \pubyear{2010}

\maketitle

\label{firstpage}

\begin{abstract}
The survey phase of the {\it Kepler Mission} includes a number of hot subdwarf B (sdB) stars to search
for nonradial pulsations.  We present our analysis of two sdB stars that are found to be $g$-mode pulsators
of the V1093\,Her class.  These two stars also display the distinct irradiation effect typical of sdB stars with a close M-dwarf companion with orbital periods of less than half a day.  Because the orbital period is so short,  
the stars should be in synchronous rotation, and if so, the rotation period should imprint itself on
the multiplet structure of the pulsations.  However, we do not find clear evidence for such rotational splitting.  Though the stars do show some frequency spacings that are consistent with synchronous rotation, they also display multiplets with splittings that are much smaller.  Longer-duration time series photometry will be needed to determine if those small splittings are in fact rotational splitting, or caused by slow amplitude or phase modulation.  Further data should also improve the signal-to-noise, perhaps revealing lower amplitude periodicities that 
could
confirm the expectation of synchronous rotation. The pulsation periods seen in these stars show period spacings that are suggestive of high-overtone $g-$mode pulsations.
\end{abstract}


\section{Introduction}

Subdwarf B (sdB) stars are horizontal-branch stars with masses
$\approx$0.47\,M$_\odot$, thin ($< 10^{-3}$\,M$_\odot$) hydrogen surface layers,
and effective temperatures from $22\,000$ to $40\,000$\,K
\citep{heber,saf94}.
They represent the aftermath of the core helium flash, and as such are important
stars for understanding this rapid phase of stellar evolution.
Some of these stars are nonradial pulsators, and we can exploit this by using
the tools of asteroseismology to probe their
interior structure, as well as determine accurate  estimates of their total mass, shell
mass, luminosity, internal rotation, and the extent of radiative levitation and gravitational diffusion. 

Pulsating sdB stars come in two varieties. The first type to be found was the
short-period V361\,Hya stars \citep{kilk97}.  They are typically referred to as
sdBV stars. Their pulsations are $p$-modes, with periods that range from two to four minutes, 
but they can range
up to  ten minutes; pulsation amplitudes are typically near 1 per cent of their mean brightness
\citep{Kilk07, reed07b}. The second variety are longer-period $g-$mode pulsators, first discovered
in 2003 \citep{grn03,mdr2004} and known as V1093\,Her (or PG\,1716) stars. 
They are commonly referred to as long-period sdBV (lpsdBV) stars.
Their periods range from 45\,min to more than 2\,hr with amplitudes typically
$<$\,0.1 per cent though amplitudes up to 0.5 per cent have been observed \citep{RoyJENAM}.
They also are all multimode pulsators.  Examples of the large number of pulsation modes determined using ground-based observations are PG\,1338+481, with
13 periodicities \citep{rand06b}, and PG\,1627+017 with 23 periodicities \citep{rand06a}. There is also a group of hybrid stars
that pulsate with both short and long periods \citep{oreiro, schuh06}.  

Those stars that are $p$-mode pulsators allow us to probe the outer layers of these stars to below
the transition from hydrogen to helium.  For understanding their cores - and directly probing the 
regions where helium is undergoing fusion - we need accurate measurements of $g$-modes \citep{charp00}.
The observational challenges in accurately identifying the pulsation modes are significant.
This usually requires extensive
photometric campaigns, preferably at several sites spaced in
longitude to reduce day/night aliasing. However, {\it Kepler} provides long-duration 
continuous, homogenous, evenly spaced time-series photometry, making it an
ideal instrument for asteroseismology.

The striking advantage of \emph{Kepler} data is the many improvements
over ground-based data, particularly for long-period sdBV variability.
From the ground, one can only observe a few pulsation cycles during
nighttime hours.  If multisite observations are obtained, then
differing instrument sensitivities become an issue, and because of weather,
gaps will doubtlessly appear in the data. Additionally, ground-based
observations have to contend with atmospheric transparency variations
that can often act on the same time scales as the pulsations in the stars. In contrast, 
\emph{Kepler} data are optimal for these types of pulsators
in that nearly gap-free data are obtained at a roughly constant cadence
and with no atmospheric issues. 
The combined advantage in signal--to--noise and coverage allows us to detect
many more pulsation frequencies than Earth-bound telescopes can obtain,
even with significant effort.

Because of their asteroseismic potential, sdB stars (and white dwarfs) are targets that form a key component in the {\it Kepler Mission}'s asteroseismic investigation. The \emph{Kepler Mission} science
goals, mission design, and overall performance are reviewed
by \citet{bor10} and \citet{koch10}. Asteroseismological analysis for the \emph{Kepler Mission}
is being conducted through the {\it Kepler} Asteroseismic Science Consortium
(KASC)\footnote{At the time of this work, the Kepler Asteroseismic Investigation, which manages the KASC, was led by
R. Gilliland, T. Brown, J. Christensen-Dalsgaard, and H. Kjeldsen.} -- see \citet{Gill10}. Compact pulsators, including sdB stars and white dwarfs, are the responsibility of the Compact Pulsators Working Group (WG11), which is overseeing the analysis of the survey phase data on pulsating hot subdwarfs.  WG11 target selection, spectroscopic properties of the targets, and overall results of the search for pulsators are described in \citet[Paper I]{roy10P1}. The first short period  sdBV star found by {\it Kepler} is presented in  \citet[Paper II]{sdBVP1}.  Long-period sdB pulsators found in the first part of the survey phase are described in \citet[Paper III]{lpsdBVP1} with detailed modelling of one of those stars in \citet[Paper IV]{vang10}. 

One of the lingering mysteries in stellar astrophysics is the mechanism by which sdB stars form.
There are many of these stars, all of which have very thin surface hydrogen layers.
Their relatively high effective temperatures signify surface hydrogen layers ranging from $10^{-3}$ to 
$10^{-5}$M$_{\odot}$.  One of the leading formation scenarios relies on binary mass transfer as the
sdB progenitor climbs the first giant branch \citep{han02, han03} and, in fact, many sdB stars (pulsators and non-pulsators)
show evidence of close companions
\citep{Maxted01,
luisa03,
napi04,
Heber09}.
In several cases, pulsating sdB stars are in close binaries with orbital periods of a few hours to 1.2 days or so \citep{kilk98, bill00, simon, geier10}.  
Because the orbital period is so short, the stars should
be in synchronous rotation \citep{zahn75,zahn77,GoldNic89}.  Thus these 
systems can help test ideas of tidal synchronization through spectroscopy \citep{geier10} and photometry \citep{bloem10}.
Asteroseismic studies suggest that the sdB component of the binaries Feige 48 \citep[$P$\,$\approx$\,9\,h]{vang08} and PG\,1336-018 \citep[$P$\,=\,2.4\,h]{charp08} are indeed in synchronous rotation.  In Feige\,48, the companion is probably a white dwarf \citep{simon}, while in PG\,1336-018 the companion is an M5 star \citep{vuc1336}.

In this paper we examine two long-period sdB pulsators
discovered during the survey phase of the \emph{Kepler Mission}: KIC\,02991403 and KIC\,11179657. Both of these stars show evidence for low-amplitude variation revealing a close binary companion with an orbital period of 9-10\,hr.  The photometric variations  suggest a lower-mass companion displaying the reflection effect.  The parameters of the binary systems (beyond the orbital period and amplitude of photometric variation) are at the moment not known, so in this paper we concentrate on analysis of the pulsations and defer a detailed discussion of the binary properties to a later paper.  We note that another complex pulsating sdB star in a close binary was uncovered in the survey phase, with preliminary analysis presented in \citet{ostensen2010b}.

\section{Observations}
This paper describes data obtained by \emph{Kepler} during
2009.  The data released to the KASC Compact Stars Working Group (WG11) 
are short cadence (SC) data with an average cadence of 58.85\,s.  This
pipeline provides time series of ``raw'' flux values for each target, and
``corrected'' fluxes based on preliminary estimates for contamination by
nearby stars.  Because the contamination correction
will be improved with further upgrades to the data pipeline, in this paper we work
with the raw fluxes provided to the KASC.  For analysis of the flux variations, we
consider fractional variations away from the mean flux (i.e., differential intensity
$\Delta I / I$).   Amplitudes are
given as milli-modulation amplitudes (mma), with 10\,mma
corresponding to 1.0 per cent.

While the data from {\it Kepler} are nearly continuous, there are occasional
gaps in the data as the result of safing events and other brief anomalies.
The window function of the power spectrum is essentially a sinc$^2$ function with a width equal to the inverse of the run length, since the duty cycle of the reduced  
light curves is so high, and the outlying points are scattered 
more-or-less randomly through the light curves.
The time series data provided from the SC data frequently show instrumental
artefacts that are at multiples of the long-cadence (LC) readout rate of 1/30 the 
SC rate.  This produces peaks in the temporal spectrum at $n\cdot 566.44\,\mu$Hz,
and experience shows that the largest-amplitude artefact peaks are in the 4000 to 7000
$\mu$Hz region \citep{Gill10SC}.
All variation frequencies in these data are well short of $566.4\,\mu$Hz 
and so we are not influenced by this artefact.

The two targets were selected using the process described in Paper I; further details about
these two targets can be found in that paper.  These two stars are the faintest of the eight pulsating 
sdB stars uncovered in the first half of the survey phase.

\paragraph*{KIC\,02991403} This target was selected based on photometry from the {\it SDSS/SEGUE} survey \citep{sdss, segue}.
With a {\it Kepler} magnitude $Kp$ of 17.136, this star
 is a spectroscopically confirmed sdB star with
$T_{\rm eff}= 27,300 \pm 200$\,K and $\log g = 5.43\pm0.03$. 
The spectroscopic parameters of this star, and those of KIC 11179657 quoted below, were estimated from spectra obtained with the ISIS spectrograph on the William Herschel Telescope, having a resolution of $R$\,$\sim$\,1600.
For details on the spectroscopic determination of these quantities, see Paper I.
This star lies within the long-period sdB instability region in the H--R diagram.
Time-series data were obtained during the first quarter year of {\it Kepler} science operation (Q1) over
a 34\,d span between BJD~2454964
and 2454997 (2009 May 12 and 2009 June 14). This provides a formal frequency
resolution ($1/T$) of $0.35\mu$Hz. 
Since {\em Kepler} data prior to Q2 were delivered without a barycentric
correction, we applied the BJD correction using the documented procedures provided by the {\em Kepler} program
\footnote{J.~Van~Cleve (ed.) Kepler Data Release Notes 2, available at:  http://archive.stsci.edu/kepler/data\_release.html}.
The {\it Kepler} Input Catalog (KIC) lists a contamination factor of
0.601 for this star.  As mentioned earlier, we used the raw flux for KIC\,02991403.
When contamination effects are applied,
the amplitude will be higher than the raw amplitude by 
a factor of $(1-c)^{-1}$, where c is the contamination factor.
The KIC estimate of $c$ is expected to be improved upon in subsequent analyses and 
data releases.
The raw data contain a significant number of outlier points - we removed these points by first
boxcar filtering the data to remove the orbital effects, and then identifying, for later removal, points that were in
excess of 4 times the RMS deviation of points from the mean.  This removed 1177 points from
the raw light curve (roughly 2 per cent of the total number of data points),
leaving 47,939 data points for analysis.  Finally, a slow decrease
in overall flux was present in the Q1 data on this star, so we fitted and removed a 2nd-order polynomial
trend from the data.

\paragraph*{KIC\,11179657} 

We identified this star as a potential target based on photometry from the AIS survey of the {\it GALEX} satellite \citep{galex}. This star has a {\it Kepler} magnitude $Kp$  of 17.065.  Spectroscopic determination (Paper I) provides an 
effective temperature of $26,000 \pm 800$\,K and $\log g=5.14\pm0.13$.  This places it within the long-period
sdBV region of the H--R diagram, but at the low-gravity end. {\it Kepler} data were obtained during the second quarter
during a 27\,d span between BJD~2455064
and 2455091 (2009 August 20 to  2009 September 16). This provides a $1/T$ frequency
resolution of 0.43\,$\mu$Hz.
A total of 39,810 science
images were scheduled. However because of safing events, loss of fine
guidance, and other small glitches 1,492 images were either not 
obtained or were contaminated in some way.   As for KIC\,02991403, 
we de-trended the slow variation in the mean level of the data, and removed
4-$\sigma$ outliers, resulting in a total of 38,272 data points in the reduced
light curve.  The KIC lists a contamination factor of 0.129, thus the amplitudes we
determine in the time-series analysis using the raw fluxes are closer to the
true amplitude in the star than for KIC\,02991403.

\section{Analysis}

Our analysis proceeded in a straightforward way as has been used in prior papers in
this series \citep{sdBVP1, lpsdBVP1}. We calculated a
temporal spectrum (a Fourier transform; FT) which provides initial estimates of
pulsation frequencies and amplitudes. We simultaneously fit and
prewhitened the data for each of these frequencies using
non-linear least-squares (NLLS), beginning with the
highest amplitude frequency until we reached the
4-$\sigma$ detection limit.  We determined $\sigma$ by finding the mean
value of the temporal spectrum in the frequency range of interest, excluding 
obvious periodicities.  We begin here with a discussion of the binary light curves, and the assumed impact
that rotation should have on the pulsations as we search for evidence of rotational splitting at the orbital / co-rotation frequency.  We follow with analysis of the higher-frequency pulsational periodicities in the two stars.  

\subsection{Binary light curves}

\begin{figure*}
\begin{minipage}{126mm}
\includegraphics[height=126mm,angle=-90]{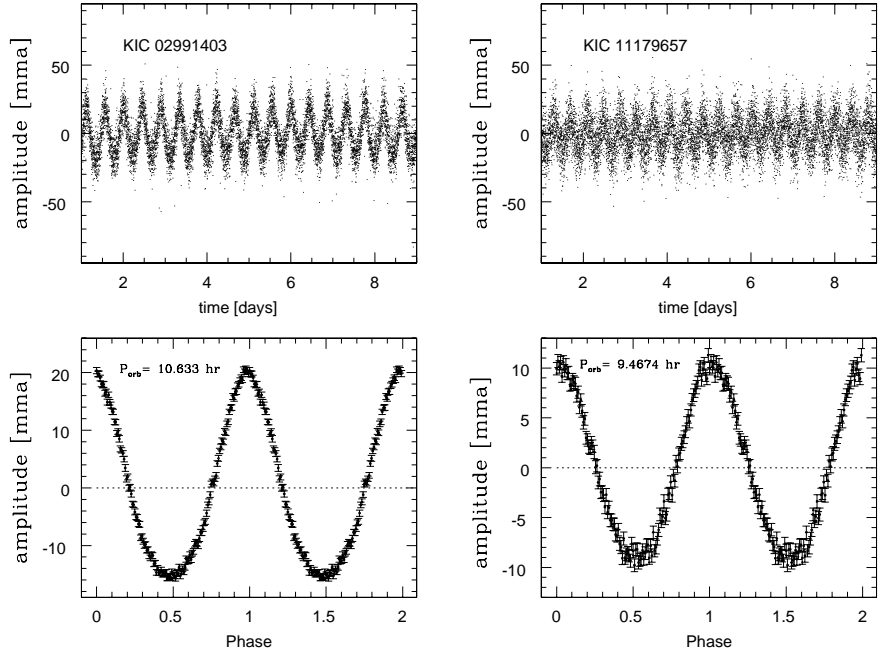}
\caption{The light curves of two sdB binary stars.  For each star, the top panel shows an 8-day sample of the SC light curve.  In the bottom panels, the time-series data (with the pulsations filtered out) are folded at the
indicated binary period, and binned into 300~s phase bins.  Error bars represent the scatter about the mean
for each phase bin and include, in part, the pulsations.  The left panel shows data for KIC\,02991403, and the
right panels show KIC\,11179657.} \label{fig01}
\end{minipage}
\end{figure*}

Both of these stars show variations with periods of about 10\,h.  KIC\,02991403 and, to a lesser extent, KIC\,11179657, show equal amplitude, peaked maxima and shallow minima, typical of the irradiation effect produced when an sdB star heats up the face of an M-dwarf orbiting the subdwarf (with its rotation period likely synchronized to the orbital period).  These effects produce light curves with a dominant peak at
the orbital frequency, and a smaller peak at the first harmonic.
The bottom panels of Figure \ref{fig01} show the binned light curves for the two stars. 
The specifics of the binary properties, including spectroscopic observations,
will appear in a subsequent publication.

No eclipses are evident in either light curve, and the amplitude is quite low compared to other known sdB+dM binaries.  For comparison, the amplitude of the reflection effect in the sdB+dM binary HE~0230-4323 \citep{koen07,Kilk10} is about 4 per cent in V, and the orbital period is 10.8\,h.  The eclipsing sdb+dM binary PG~1336-018 (orbital period of 2.4\,hr) is almost 10 per cent \citep{ kilk98}. Since PG~1336-018 is an eclipsing system, the inclination must be close to 90$^\circ$ and so represents the intrinsic reflection effect amplitude. On the other hand, if the amplitude of the effect drops with the square of the orbital separation, then the intrinsic amplitude of the reflection effect is approximately 6.3 times smaller, or 16\,mma.  If that is the case, then these systems may be closer to edge-on, since the amplitude of the reflection effect that we see is then comparable to what we would expect.  In the opposite extreme, if the intrinsic amplitude is similar for these somewhat longer-period binaries, then the low amplitudes we see (down by a factor of 10) could indicate a much smaller inclination of $5^\circ$ or so.  From the information we currently have about these stars, it is difficult to say what, if any, constraints the binary light curve places on the orbital inclination.  Time-series spectroscopy will be required to constrain the companion type and inclination angle further.

In the left panels of Figure \ref{fig02} and Figure \ref{fig03} we show the temporal spectrum
of the frequency region containing the binary period and first harmonic.
The right panels in these figures show the area of the temporal spectrum where pulsations appear.

%fig01

For this analysis, we are primarily
interested in how the binarity impacts the variations attributed to
pulsations. For close binaries such as these, one expects that the stars will be in synchronous rotation - that is, the rotation  frequencies of the stars should match the orbital frequency of the binary ($\Omega = f_{\rm orb}$, where $\Omega$ is the rotation frequency and $f_{\rm orb}$ is the orbital frequency), and that the rotation and orbital axes are aligned.
For nonradial pulsation, rotation should cause a shift in the observed oscillation frequencies for modes with nonzero values of the azimuthal quantum number $m$.  In particular, for slow rotation ($\Omega \ll \nu$, where  $\nu$ is the pulsation frequency) it is well known that:
\begin{equation}
\nu_{n,l,m} = \nu_{n,l,0} + m \Omega (1-C_{n,l})
\end{equation}
\citep{ledoux}.  Here, the indices $n$, $l$, and $m$ correspond to the radial order $n$ and the indices of the spherical harmonic functions $Y_l^m$. The quantity $C_{n,l}$ describes the contribution to the splitting caused by the Coriolis force, and is a function of the eigenfunction for the appropriate nonradial mode.
Rotation, therefore, can  break the degenerate $m$ frequencies into
their $2l +1$ components separated approximately by the rotation 
frequency. As such, multiplets can be used to associate pulsation
modes (described by $n$, $l$, and $m$) with observed frequencies.
Seismic models of stars provide the theoretical oscillation frequency spectrum, which can
be quite rich; two similar models can show similar frequencies, but with different values
for $n$, $l$, and $m$.  Therefore mode identifications are needed to
constrain stellar models, which in turn are used
to infer internal conditions of the stars themselves.

For the long-period pulsations seen in these stars, we expect that they are high-order $g$-modes.  A useful approximation for these modes, where the horizontal motions are significantly larger than the radial motions, is that
\begin{equation}
C_{n,l} \lesssim \frac{1}{l(l+1)}.
\end{equation}
 Consulting evolutionary models in the relevant mass, $T_{\rm eff}$ and $\log g$ ranges from \citet{kaw05} and \citet{char02}, we find that $C_{n,l}$ ranges from 0.465 to 0.496 for $l=1$ $g$-modes in the observed period range (see below) and between 0.158 to 0.165 for $l=2$ modes. So the above approximation appears reasonable for these stars.  We note that if there is some tidal deformation of the sdB stars in these binaries, that there may be additional frequency shifts associated with that change in stellar shape (i.e. one would expect equally spaced multiplets for all modes with the same spacing if the tidal axis coincides with the pulsation axis).  For this initial analysis, we assume that the pulsators are spherically symmetric.
 
While rotation can lift the $m$ degeneracy and produce $2 l + 1$ peaks for each value of $n$ and $l$, the viewing geometry (as well as the intrinsic pulsation amplitude) will alter the observed amplitudes. The geometric effects, at least, can be computed in a straightforward way \citep[see, for example,][]{pes85}.  
For these stars, which show orbital effects in the light curve, we expect that the stars are in synchronous rotation at the orbital period.  
% If we are viewing them at relatively high inclination, the sectoral modes ($m=\pm l$ 
% are favored for $l=1$ modes, and for $l=2$ 
% modes, the sectoral modes with the central $m=0$ mode are favored.  
Thus, $l=1$ modes are likely to appear as multiplets  split by 0.51 to 0.54 $f_{\rm orb}$ with up to 3 peaks for each overtone $n$. For $l=2$ we should look for a multiplet splitting of about $0.84 f_{\rm orb}$, with up to 5 peaks for each overtone.

\subsection{KIC\,02991403}
\begin{figure*}
\begin{minipage}{135mm}
\includegraphics[height=135mm,angle=-90]{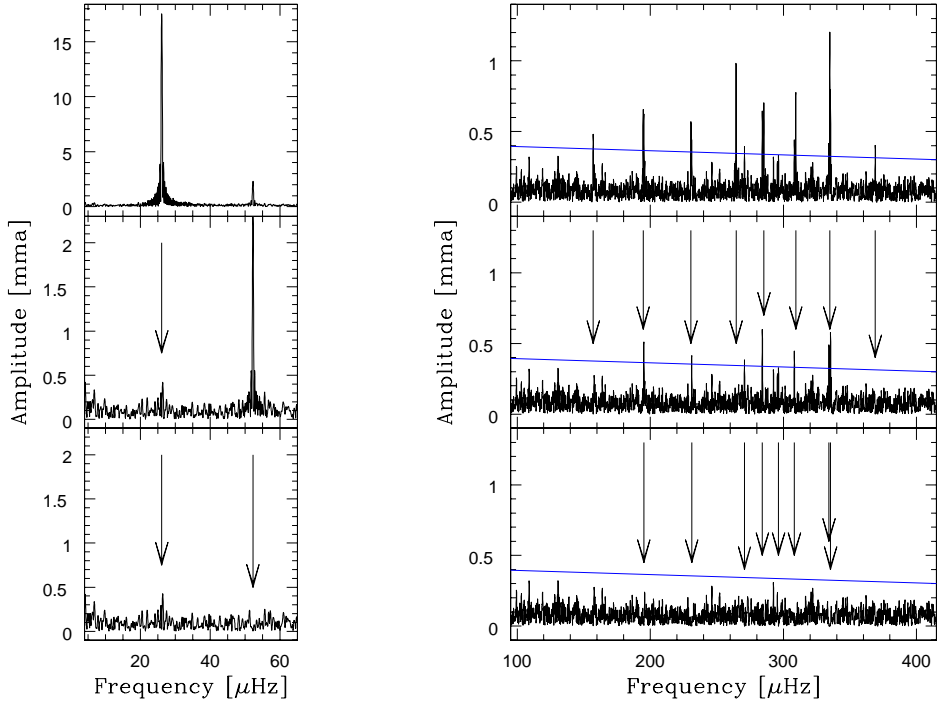}
\caption{Temporal spectra of KIC\,02991403.  The left panels show the peaks attributed to binary effects.  The top panel shows the temporal spectrum of the data with the clear signature of the orbit.  The middle panel has had that principal periodicity removed, exposing the first harmonic of the orbital period.  The bottom panel shows the small residual spectral features after removal of the fundamental and first harmonic.  The panels on the right show the peaks attributed to long-period pulsations in the sdB star.  Again, the top panel shows the temporal spectrum of the observed light curve; the middle panel shows the residuals after removal of the highest eight peaks, and the bottom panel shows residuals following removal of all peaks in Table \ref{tab01}.  The blue (sloping horizontal) line denotes the 4-$\sigma$ detection threshold.} \label{fig02}
\end{minipage}
\end{figure*}

The temporal spectrum of KIC\,02991403, prior to prewhitening, is shown in Figure \ref{fig02}, in the top panels. There are several artefact frequencies in the 4500-5500 $\mu$Hz range, but  most of the peaks in the FT of this star occur in the 100-400 $\mu$Hz range.  We determine a  4-$\sigma$  detection limit using apparently variation-free frequency regions above and below this range.   Between 1 and 150\,$\mu$Hz, the 4-$\sigma$ limit is 0.40\,mma, and between 400 and 550\,$\mu$Hz we find a value for the 4-$\sigma$ limit of 0.31\,mma.  This detection limit is shown as a blue (mostly horizontal) line in Figure \ref{fig02}.

The middle and bottom panels of Figure~\ref{fig02} show the detection limit
and the prewhitening process for KIC2991403. The highest amplitude
variation has a frequency near $26.1\,\mu$Hz and a period near
0.44\,d. This frequency and its associated overtone are attributed
to  binarity and shown in the left panels of Figure~\ref{fig02}.
We detected
16 other frequencies that we attribute to long-period $g-$mode pulsations.
These are listed in Table~\ref{tab01} along with their amplitudes and
periods and
indicated in Figure~\ref{fig02} with arrows. No peaks remain in the FT above the
detection limit after removal of these periodicities.

% fig02

Table~\ref{tab01} lists the resulting periodicities.  Of the 16 pulsation frequencies
in the table, there are 5 sets of closely spaced multiplets.
We list these close spacings, along with the spacings expected for synchronous (and uniform) rotation
in Table \ref{tab02}. The smallest of these spacings, 0.47\,$\mu$Hz, is a bit larger than the formal frequency
resolution of 0.35\,$\mu$Hz. The other close spacings have values of approximately 0.55 to 0.58\,$\mu$Hz or multiples thereof. If these small spacings are interpreted as a rotational splitting
of $l=1$ modes, then the implied rotation period (using Equation 1 and 2) is 10.4\,d.  If they are $l=2$ sectoral $m=\pm2$ modes, then the rotation period is 34\,d.  In either case, if this splitting is indeed caused by
rotation, then this star is far from
synchronous rotation.  To obtain a more precise estimate of these
spacings (or to see if more periodicities are present than suspected) will require a much longer
run on this star. 

% Table 1
\begin{table}
%\begin{minipage}{126mm}
\caption{Frequencies, periods, and average amplitudes for KIC\,02991403.
The binary period and first harmonic are included as well. 
Formal least-squares errors are in parentheses in this and subsequent tables.\label{tab01}}
\begin{tabular}{lccc}
\hline
ID & Frequency [$\mu$Hz] & Period  [s] & Amplitude [mma] \\
\hline 
\multicolumn{4}{c}{Binary Period and first harmonic}  \\
$f_{\rm orb}$ & 26.1240 (0.0011) & 38278.9 (1.6) & 17.55 (0.10) \\ 
$2f_{\rm orb}$ & 52.2466 (0.0084) & 19140.0 (3.1) &  2.33 (0.10) \\
\multicolumn{4}{c}{Pulsation Frequencies}  \\
f1  & 157.101 (0.025) & 6365.35 (1.01) &  0.484 (0.063)     \\
f2  & 194.675 (0.028) & 5136.77 (0.75) &  0.556 (0.064)    \\
f3  & 195.144 (0.031) & 5124.41 (0.81) &  0.513 (0.064)   \\
f4  & 230.552 (0.022) & 4337.42 (0.42) &  0.553 (0.063)  \\
f5  & 231.115 (0.030) & 4326.84 (0.55) &  0.416 (0.065)  \\
f6  & 264.437 (0.012) & 3781.62 (0.18) &  0.980 (0.065)  \\
f7  & 270.739 (0.032) & 3693.59 (0.43) &  0.382 (0.066)  \\
f8  & 284.149 (0.020) & 3519.28 (0.25) &  0.601 (0.063)  \\
f9  & 285.309 (0.019) & 3504.97 (0.23) &  0.661 (0.064)  \\
f10 & 296.303 (0.036) & 3374.93 (0.41) &  0.332 (0.063)  \\
f11 & 308.188 (0.028) & 3244.77 (0.30) &  0.434 (0.063) \\
f12 & 309.268 (0.016) & 3233.44 (0.16) &  0.783 (0.065)   \\
f13 & 334.233 (0.030) & 2991.93 (0.27) &  0.450 (0.063)   \\
f14 & 334.818 (0.013) & 2986.70 (0.11) &  1.144 (0.063)  \\
f15 & 335.390 (0.026) & 2981.60 (0.23) &  0.544 (0.064) \\
f16 & 369.013 (0.030) & 2709.93 (0.22) &  0.402 (0.067) \\
\hline
\end{tabular}
%\end{minipage}
\end{table}

On the other hand, the expected spacing for consecutive $m$ modes in an $l=1$ multiplet is $13.60\pm0.34$\,$\mu$Hz, where the uncertainty comes from the range of $C_{n,l}$ for relevant models.  Table~\ref{tab02} shows two spacings that are consistent with $l=1$ rotational splitting at the orbital frequency.  The $\Delta m=2$, $l=1$ spacing involves the closely spaced pair f14,15 on the high frequency side. Complicating this picture further is the observation that the separation between f15 and f12 is (to well within measurement error) precisely equal to the orbital frequency.  This suggests some geometric component to the overall frequency structure.

% Table 2
\begin{table}
\caption{Frequency spacings in KIC\,02991403.\label{tab02}}
\begin{tabular}{lcc}

\hline
IDs & Spacing [$\mu$Hz] & Comment\\
\hline
\multicolumn{3}{c} {Exact orbital frequency spacings}\\
f15 - f12 & 26.122 (0.031) & $f_{\rm orb}-0.002\,\mu{\rm Hz}$ \\
f13 - f11 &  26.046 (0.043) &  $f_{\rm orb} - 0.079\,\mu{\rm Hz}$ \\
 & & \\
\multicolumn{3}{c}{close spacings near run resolution}  \\
f3 - f2  & 0.469 (0.042) & \\
f5 - f4  & 0.548 (0.037) \\
f9 - f8     & 1.160 (0.028) &  = $2 \times 0.580$ \\
f12 - f11   & 1.080 (0.033) & = $2 \times 0.540$ \\
 & & \\
\multicolumn{2}{c}{closely spaced triplet}  \\
f14 - f13  & 0.585 (0.033) & \\
f15 - f14   & 0.572 (0.029) \\
 & & \\
\multicolumn{3}{c}{$l=1$ splitting = $13.60\pm0.34\,\mu$Hz}\\
f14 - f11 & 26.630 (0.031) & $\Delta m = 2$ \\
f15 - f11 & 27.202 (0.038) & $\Delta m = 2$ \\
f8 - f7 & 13.410 (0.038) & $\Delta m = 1$ \\
& & \\
\multicolumn{3}{c}{$l=2$ splitting = $21.91\pm0.10\,\mu$Hz}\\
f10 - f4 & 65.750 (0.036) & $\Delta m = 3$ \\
f11 - f6 & 43.751 (0.030) & $\Delta m = 2$ \\
& & \\
\multicolumn{3}{c}{$l=3$ splitting = $24.06\pm0.10\,\mu$Hz}\\
f12 - f9 & 23.959 (0.025) & $\Delta m = 1$\\
f11 - f8 & 24.039 (0.034) & $\Delta m = 1$\\
\hline
\end{tabular}
\end{table}

% % fig03
% \begin{figure}
% \psfig{figure=kpr2991403mult.ps,angle=-90,width=\textwidth}
% \caption{Schematic temporal spectrum indicating multiplets for 
% KIC2991403.} \label{fig03}
% \end{figure}

With many periodicities present in a narrow frequency range, the question arises as to
whether the number of pulsation modes available to the star is sufficient to explain the
number of periodicities seen.  However, unlike the short-period sdB pulsators, this is not an 
issue for lpsdBV pulsators \citep[e.g. Fig. 7 in][]{mdr2004}.
Stellar models show many closely spaced frequencies in this region
\citep{font06,jeff07,vang10}.  Still, pulsation models have found it easier to drive
high-degree modes and since the detected amplitudes continue down to the 
detection limit, and there are likely pulsations below our limit, it
may be that high-degree modes are present. \emph{Kepler} can test this
critical model assumption in that with more data and a lower detection
threshold, it may be possible to determine if $l>3$ are indeed
present. As $l\geq 3$ have a large degree of
geometric cancellation \citep{charp05a,me1}, if their amplitudes are
intrinsically similar to low-degree modes, then their observed amplitudes
must be significantly reduced.

\subsection{KIC\,11179657}

\begin{figure*}
\begin{minipage}{135mm}
\includegraphics[height=135mm,angle=-90]{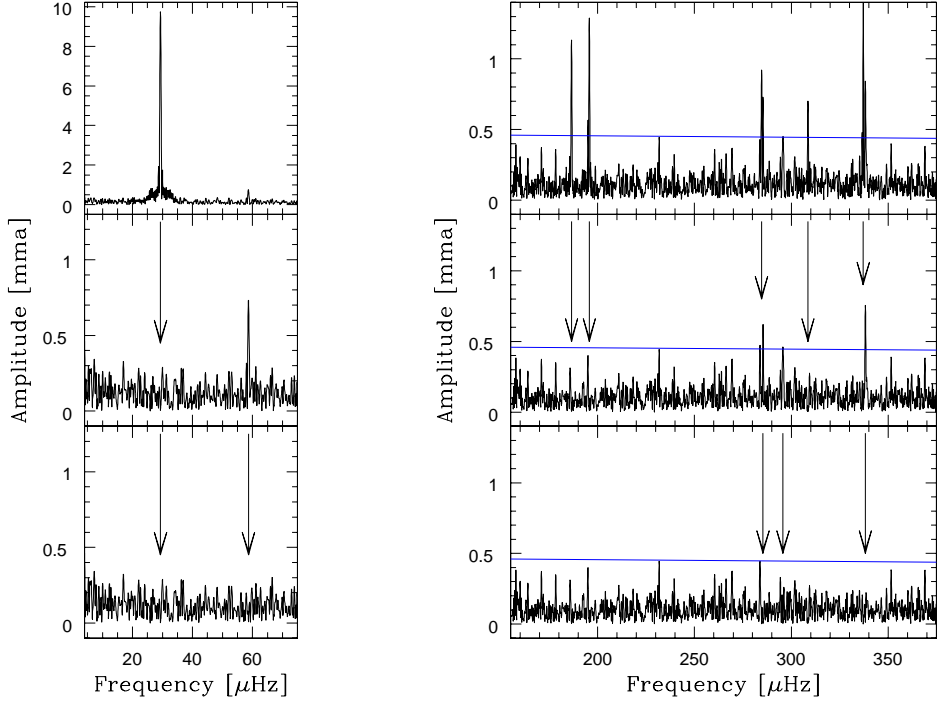}
\caption{Same as Figure \ref{fig02} but for KIC\,11179657. Arrows indicate frequencies of those periodicities that were removed at each stage of prewhitening.  The frequencies are listed in Table \ref{tab03}.  The horizontal blue line denotes the 4-$\sigma$ detection threshold.} \label{fig03}
\end{minipage}
\end{figure*}

As for KIC\,02991403, we determined the 4-$\sigma$  detection limit
for KIC\,11179657 using apparently variation-free frequency regions on either side
of the range where obvious periodicities are present,
and interpolating in between. On the low frequency side, near 100\,$\mu$Hz, the 4-$\sigma$ level is 0.466\,mma, and
the noise level falls slightly to 0.431\,mma near 400\,$\mu$Hz.
The blue lines in Fig~\ref{fig03} show the detection limit
and the prewhitening process for KIC\,11179657. Vertical arrows denote periodicities removed from
the temporal spectrum at each stage. The lowest-frequency variation
has a frequency near $29.4\,\mu$Hz and a period near
0.39\,d. This frequency and its associated harmonic are attributed
to  binarity and shown in the left panels of Fig~\ref{fig03}.
We detected
8 other frequencies with confidence, and as in the previous section we attribute these to long-period 
$g-$mode pulsations.
The identified periodicities are listed in Table~\ref{tab03}. Prewhitening effectively removes nearly all power near
the detection limit.

% fig03
% fig02

Two peaks in the bottom panel of Fig.~\ref{fig03} indicate other periodicities just below the
4-$\sigma$ detection limit, but likely intrinsic to the star.  We list these as tentative peaks in Table~\ref{tab03}, 
but for this paper we adhere to the 4-$\sigma$ limit in identifying secure periodicities. There may indeed be other periodicities present, but we defer identification
of further peaks to upcoming studies with longer time-series data, which we anticipate obtaining
in the future with \emph{Kepler}. Again, this
is likely as \emph{Kepler} continues to monitor this star;  
longer runs may allow us to detect high-degree modes
which suffer large geometric cancellation.

% Table 3
\begin{table}
%\begin{minipage}{126mm}

\caption{Frequencies, periods, and average amplitudes for KIC\,11179657.
The binary period and harmonic are included as well. 
Formal least-squares errors are in parentheses.\label{tab03}}
\begin{tabular}{lccc}
\hline
ID & Frequency [$\mu$Hz] & Period [s] & Amplitude [mma] \\
\hline

\multicolumn{4}{c}{Binary Period and first harmonic}  \\

$f_{orb}$ & 29.3403 (0.0024) & 34082.8 (2.8) & 9.74 (0.10) \\
$2 f_{orb}$ & 58.671 (0.032) & 17044.3 (9.4) & 0.74 (0.10) \\
\multicolumn{4}{c}{Pulsation Frequencies}  \\

f1 & 186.517 (0.018) & 5361.43 (0.53) &  1.115 (0.088)  \\
f2 & 195.700 (0.016) & 5109.87 (0.42) &  1.275 (0.086)  \\
f3 & 284.656 (0.025) & 3513.01 (0.31) &  0.864 (0.088)  \\
f4 & 285.393 (0.034) & 3503.94 (0.42) &  0.626 (0.086)  \\
f5 & 295.690 (0.044) & 3381.92 (0.50) &  0.468 (0.086)  \\
f6 & 308.597 (0.030) & 3240.47 (0.31) &  0.695 (0.086)  \\
f7 & 337.167 (0.014) & 2965.89 (0.13) &  1.446 (0.086)  \\
f8 & 338.256 (0.028) & 2956.34 (0.24) &   0.755 (0.086) \\
 & \\
\multicolumn{4}{c}{Additional periodicities near 4-$\sigma$ cutoff}  \\
 
s1    & 231.634 (0.037) & 4317.15 (0.70) & 0.447 (0.086) \\
s2    & 283.843 (0.044) & 3523.07 (0.54) & 0.451 (0.086) \\
\hline
\end{tabular}
%\end{minipage}
\end{table}

In Table~\ref{tab04} we list spacings that correspond to multiples of the orbital frequency, close splittings, and spacings expected for synchronous rotation.
A single splitting near the orbital frequency is detected. This
is between f6 and f8, with a spacing of 29.66\,$\mu$Hz, which is, within the formal frequency resolution of the data, equal to the orbital frequency. 
Among the
remaining periodicities, there is the suggestion of a closely spaced triplet (s2, f3, f4) with an average spacing of 0.775\,$\mu$Hz, with one of the three components being the tentative peak s2.  If this is indeed a rotationally split triplet, then  KIC\,11179657 rotates with a period of 7.62\,d (if $l=1$).  This is long compared to the orbital period. 

We see no spacings in the current set of frequencies that could be an $l=1$ rotationally split multiplet at the synchronous rotation period.  There are frequency differences that are at the expected spacing for $l=2$ multiplets.  Interestingly, two of the closely spaced frequencies, f3 and f4, are (within the errors) at 4 times the expected $l=2$ spacing from f1.  Again, though, like KIC\,02991403, the high frequency component is split by a smaller value (the close spacing noted above).
Considering these complications and the limits of signal-to-noise and frequency resolution, no firm conclusions about rotation in KIC\,11179657 can be drawn at this point.

% Table 4
\begin{table}
\caption{Frequency spacings in KIC\,11179657.\label{tab04}}

\begin{tabular}{lcc}
\hline
IDs & spacing [$\mu$Hz] & Comment\\
\hline
\multicolumn{3}{c} {Exact orbital frequency spacings}\\
f8 - f6 & 29.659 (0.041) & $f_{\rm orb} + 0.319\,\mu{\rm Hz}$ \\
 & & \\
\multicolumn{3}{c}{close spacings near run resolution}  \\
f3 - s2  & 0.813 (0.050) & \\
f4 - f3   & 0.737 (0.042) & \\
f8 - f7  & 1.087 (0.031) & \\
 & & \\
\multicolumn{3}{c}{$l=1$ splitting = $15.27\pm0.38\,\mu$Hz}\\
none & \\
& & \\
\multicolumn{3}{c}{$l=2$ splitting = $24.61\pm0.12\,\mu$Hz}\\
f6 - s2  & 24.754 (0.053) & $\Delta m = 1$ \\
f3 - f1 & 98.14 (0.024) & $\Delta m = 4$ \\
f4 - f1 & 98.88 (0.038) & $\Delta m = 4$ \\
 & & \\
\multicolumn{3}{c}{$l=3$ splitting = $27.05\pm0.12\,\mu$Hz}\\
f4 - s1 & 53.76 (0.050) & $\Delta m = 2$\\
f8 - f2 & 54.41 (0.052) & $\Delta m = 2$\\
\hline

\end{tabular}
\end{table}

\section{Discussion}

\subsection{Synchronous rotation and rotational splitting}

The binary nature of these two stars, and in particular the short orbital periods, promises to provide the extra information necessary for identifying rotationally split multiplets under the assumption that the primary star -- the sdB -- is forced into synchronous rotation.  If that is the case, the discussion in Section 2 leads to the expectation that rotationally split multiplets should be seen with splittings of 13.6(21.9)\,$\mu$Hz for $l=1$($l=2$) modes in KIC\,02991403.  For KIC\,11179657, the respective expectations are 15.3\,$\mu$Hz for $l=1$ modes, and 24.6\,$\mu$Hz for $l=2$ modes.

Both stars show fine structure (that is, frequency separations that are small compared to the orbital frequency) that resembles rotational splitting for slow rotation.  Identification of almost-equally spaced triplets in these stars suggests that these splittings are ``permanent" rather than a manifestation of slow amplitude or phase modulation.   However, both stars also show larger splittings at the expected values for a synchronous rotation - for $l=1$ in the case of KIC\,02991403, and for $l=2$ in KIC\,1179657.
In both cases, the spacings imply that we are seeing sectoral $l=\pm m$ and, curiously, the higher frequency component shows fine structure (as defined above).   This splitting hierarchy has been seen in other $g$-mode pulsators - in particular, the pulsating hot white dwarf (GW~Vir) star PG~2131+066 \citep{pg2131}. 

The relatively small number of periodicities identified (so far) in KIC\,02991403 and KIC\,11179657 makes it difficult to identify multiplets with splittings compatible with the binary frequency, especially considering that the splitting depends on the value of $C_{nl}$.  We do hope that further observations will reveal more periodicities, and expect that theoretical models of this star can allow us to make more precise estimates of $C_{nl}$.  

\subsection{Asymptotics and mode identifications}

When analysing $g$-modes, we often take advantage of asymptotic behavior of the periods to try to estimate the values of $l$ and $n$.  In particular, for $g$-modes, in the absence of strong mode trapping, the periods of modes with the same $l$ and consecutive $n$ are equally spaced in period, with a fundamental period spacing $\Pi_o$ that is determined by the global structure of the star.  In particular, we have:
\begin{equation}
\Pi_{nl} = n \times \frac{\Pi_o}{\sqrt{l (l+1)}}.
\end{equation}
Thus, a sequence of $g$-modes with consecutive values of $n$ and the same value of $l$ will be spaced equally in period.  Small departures from equal spacing result from abrupt changes (with depth) in stars (e.g. density discontinuities that accompany composition gradients), and so the spacings can be  a powerful probe of the interior structure of stars. We see this behavior very clearly in pulsating white dwarfs \citep[e.g.][]{kawbra94}.  The period spacings depend on $l$ in a predictable way: modes with the same value of $n$ but with $l=1$ and $l=2$ will have periods in a ratio of approximately $\sqrt{3}$.   In those (rare) cases where sequences of $l=1$ and $l=2$ $g$-modes are confirmed (i.e. through rotationally split multiplets), we also see that the spacing ratio is indeed close to 1.7 \citep[e.g. PG~1159-035,][]{Costa08}. 

Up until now these properties of $g$-modes have not been applied to the long-period sdB stars because of the paucity of observed modes and/or secure $l$ determinations, and uncertainties in the rotational splitting (and identification of $m=0$ periods), using ground-based data.  In Paper III we reported on the pulsations of other long--period sdB pulsators that are {\it Kepler} survey targets.  In at least one case, KIC\,10670103, we do see sequences of equally spaced periods, and in fact can identify an $l=1$ sequence with a period spacing of 251\,s, and an $l=2$ sequence with a spacing of 146\,s.  The ratio is 1.72, and confirms the expectations from the asymptotic behavior of $g$-modes. 

For the two stars in this paper, are there periods that are equally spaced?  Tables \ref{tab05} and \ref{tab06} address this question by listing the periods in each star, along with the period differences between consecutive (identified) periodicities.  Though the number of periodicities is smaller than for KIC\,10670103, the periods seem to show similar spacings to that star.  The mean spacing of about 136~s in f4 through f16 in KIC\,02991403 is similar to the $l=2$ spacings in KIC\,10670103.  This spacing is inconsistent with the preliminary results based on rotational splitting, which favors an $l=1$ assignment for f11 and f14 (and f15).  In KIC\,11179657, with fewer periodicities, the period structure also shows spacings that are compatible with those seen in KIC\,02991403 and KIC\,10670103.   In summary, period spacings are insufficiently well defined to assign secure values of $l$ for these stars.  This may be because rotational splitting is present and we are not seeing $m=0$ modes.

Even though we cannot say, with confidence, that we see an equal period spacing in either of these stars, we can still use the asymptotics of $g$-modes to say something about the value of $l$. If the driving mechanism acts on a limited number of overtones ($n$), we would expect to see groups of periods in the temporal spectrum with the centroids of those groups with a period ratio of 1.73.  In KIC\,1117657, the strongest mode (f7) has a period of 2966\,s.  If that is an $l=2$ mode, the corresponding $l=1$ mode should have a period of roughly 5140\,s.  This is fairly close to f2 in that star, which is the second strongest pulsation: the ratio of f2 to f7 is 1.723.  This suggests that we are seeing $l=2$ and $l=1$ pulsations in this star.  We note that {\it Kepler} data on a single long-period sdBV star, KIC\,10670103 (Paper III), shows the same pattern of apparent $l=1$ and $l=2$ period ratios.

% Table 5
\begin{table}
\caption{Period spacings in KIC\,02991403.\label{tab05}}

\begin{tabular}{lcl}
\hline
ID & Period [s] & Difference [s] \\
\hline
f1            & 6365.35  &      \\
                             &  & $\sim 1234.8$  ($=5 \times 246.95$) \\
(f2+f3)/2     & 5130.59 &  \\ 
     & & \\
(f4+f5)/2     & 4332.13 & \\
                       &        & $\sim  550.51$  ($= 4 \times 137.63$) \\
f6            & 3781.62 & \\
                              &   &   269.50 ($= 2\times 134.75$) (f7 in between) \\
(f8+f9)/2     & 3512.12 & \\
                              &  & 137.12 \\
f10           & 3374.93  & \\
                              &  &  135.82 \\
(f11+f12)/2   & 3239.11 \\
                              &   & 252.41 (= $2\times126.21$) \\
f14             & 2986.70 & \\
                              &   &  276.77 (= $2\times138.39$) \\
f16            & 2709.93 & \\ 
\hline
\end{tabular}
\end{table}

% Table 6
\begin{table}
\caption{Period spacings in KIC\,11179657.\label{tab06}}
\begin{tabular}{lcl}
\hline
ID & Period [s] & Difference [s] \\ 
\hline
f1           & 5361.43  &      \\
                             &  & 251.56   \\
f2            & 5109.87  &    \\ 
..........& \\
s1           & 4317.15  &  \\
  & & 804.14 (=$6\times 134.02$)\\
s2,f3,f4    & 3513.01  & \\
                       &        & $\sim 131.09$ \\
f5             & 3381.92 & \\
                              &   &   141.45 \\
f6            & 3240.47 & \\
                              &  & $\sim 279.4$ (=$2\times 139.7$) \\
f7,f8             & 2956.34, 2965.9  & \\
\hline
\end{tabular}
\end{table}

Similarly, among the main periodicities in KIC\,02991403, the highest-amplitude mode has a period of 2987\,s.  Assigning that mode to $l=2$ suggests that the $l=1$ counterpart should be at a period of 5173\,s, which is very close to the doublet (f2,f3) at 5130\,s.  The ratio of f14 to f2 is 1.720.  This too is remarkably close to the asymptotic value of 1.732.  This simplistic preliminary analysis suggests that we are indeed seeing $l=2$ and $l=1$ modes.  This, in turn, means that the largest amplitude modes have $l=2$ in some cases - which is surprising since the higher-degree modes suffer more geometric cancelation across the surface.

Thus there are several lines of evidence, all circumstantial at this point, that we are seeing $l=1$ and $l=2$ $g$-modes in these two stars.    We expect to have longer runs (3 months or more) on these stars as the {\it Kepler Mission} continues, and expect that the increase in frequency precision and signal-to-noise should help us determine, with certainty, the rotational state of these stars.  Further exploration of the binary properties (i.e. a radial velocity study 
verifying the orbital and rotational signatures) should provide clues as to the properties of the (presumably dM) companions, and the nature of the binary light curves.  With this information in hand, these stars should play an important role in improving our understanding of the binary nature of sdB stars in general, and the role that binarity plays in their formation and subsequent evolution.

\section*{Acknowledgments}
For R.{\O}., C.A., and S.B., the research leading to these results has received funding from the European
Research Council under the European Community's Seventh Framework Programme
(FP7/2007--2013)/ERC grant agreement n$^\circ$227224 (PROSPERITY) and
from the
Research Council of K.U.Leuven (GOA/2008/04). ACQ is supported by the Missouri Space Grant Consortium, funded by NASA. A.B. acknowledges support from the Polish Ministry of Science (554/MOB/2009/0). 
Funding for this Discovery mission is provided by NASA's Science Mission
Directorate.
 The authors gratefully acknowledge the entire \emph{Kepler}
team, whose efforts have made these results possible.

\label{lastpage}

\end{document}